\documentclass{article}
\usepackage {amssymb}
\begin {document}
\author {Albert Schwarz
\\University of California at Davis,\\Davis,CA 95616 
   \thanks{Partially
supported by NSF
grant DMS-9801009}}
\title  {Noncommutative  Algebraic Equations and Noncommutative Eigenvalue Problem.}

\maketitle
\large
\begin {abstract}
\large
 We analyze the perturbation series for noncommutative  eigenvalue problem $AX=X\lambda$ where $\lambda$ is an element of  a
noncommutative ring,  $ A$ is a matrix and $X$ is a column vector with entries from this ring. As a corollary we obtain a
theorem about the structure of perturbation series for Tr $x^r$  where $x$ is a  solution of noncommutative  algebraic
equation  (for $r=1$ this theorem was proved by     Aschieri,   Brace,
Morariu,   and  Zumino, hep-th/0003228, and used to
study Born-Infeld lagrangian for the gauge  group $U(1)^k$ ).  
\end {abstract}

We use the term  "noncommutative algebraic equation"  for the equation of the form 
 \begin {equation}
x^n=a_1x^{n-1}+a_2x^{n-2}+...+a_n
\end {equation} 
where the coefficients $a_1,...,a_n$ and the unknown $x$ belong to an associative (but not necessarily commutative) ring
${\cal A}$. It was shown in [1] that one can prove a generalization of Vieta theorem for the roots of (1). For example, if
$x_1,...,x_n$ are roots of (1), that are independent in some sense, we have ${\rm Tr} x_1+...+{\rm Tr} x_n={\rm Tr} a_1$. Here
trace is defined as an arbitrary linear functional on ${\cal A}$ obeying ${\rm  Tr} ab={\rm Tr} ba$ for all $a,b\in {\cal A}$. 

Another proof of generalized Vieta theorem was given in [2]. This proof is based on the remark that the equation (1) is
related to "noncommutative eigenvalue problem":
\begin {equation}
\begin {array}{ccc}
a_{11}x_1+\cdots +a_{1n}x_n & = & x_1\lambda \\
 \multicolumn {3}{c} {\dotfill} \\
a_{n1}x_1+\cdots+a_{nn}x_n & = & x_n\lambda
\end {array}
\end {equation}
where the coefficients $a_{ij}$, unknowns $x_i$ and "noncommutative eigenvalue" $\lambda $ are elements of associative ring
${\cal A}$. (It is assumed  that the "noncommutative eigenvector" with
entries $x_i$ does not vanish.)  It was shown in [2] that the problem (2) arises very naturally in the analysis of linear
systems of first order differential equations in the ring ${\cal A}$ (and that (1) is related to higher order linear
differential equations in ${\cal A}$). 

Noncommutative Vieta theorem can be considered as a part of general theory
of noncommutative functions.  Noncommutative functions were studied in
important series of papers [8]-[14]. In particular, these papers contain
new proofs of noncommutative Vieta theorem   ([12],[14],[8]). More
precisely, the Gelfand- Retakh form of Vieta theorem is somewhat stronger
than the statement of [1].
  
Recently the equation (1) was studied in the framework of perturbation theory in [3]. The authors of [3] prove some unexpected
properties of perturbation series, that were conjectured for the case $n=2$ in   [4], [5]. (The equation (1) for $n=2$ appears
in the study of so called Born-Infeld lagrangian for the gauge group $U(1)^k$.)

The main goal of present letter is to state some general results about problems (1) and (2). We use these 
results to give a simple proof of the theorem of [3] and to generalize this theorem. 

Let us present the system (2) in the form 
\begin {equation}
AX=X\lambda
\end {equation}
where $A\in {\rm Mat}_{n\times n}({\cal A})$  and $X\in {\rm Mat}_{n\times 1}({\cal A})$ (i. e. $A$ is an $n\times n$ matrix,
$X$ is a column vector, both $A$ and $X$ have entries from ${\cal A}$). We will assume that ${\cal A}$ is a unital algebra
over complex  numbers and that 
$$A=\epsilon B+P\cdot 1$$ 
where  $\epsilon \in {\bf C},\ \  B\in {\rm Mat}_{n\times n}({\cal A})$ and 
 $P\in {\rm Mat}_{n\times n}({\bf C})$ is an $n\times n$ matrix with entries from ${\bf C}$. We will consider the case, when
$P$ has $n$ distinct eigenvalues $\kappa_1,...,\kappa_n\in {\bf C}$. In this case we can find $n$ solutions to the system (2)
in the framework  of perturbation theory with respect to $\epsilon$. More precisely, these solutions are formal power series 
\begin {equation}
  \lambda= \lambda ^{(0)}+\epsilon \lambda ^{(1)}+...+\epsilon ^n\lambda^{(n)}+...
\end {equation}
\begin {equation}
X=X^{(0)}+\epsilon X^{(1)}+...+\epsilon ^nX^{(n)}+...
\end {equation}
obeying (3). Using standard arguments one can show that there exist $n$ such series for $\lambda$
 $$^{(i)}\lambda =\kappa_i +{\rm higher\   order\   terms.}$$
   We denote corresponding $X$ by $^{(i)}X$ (there is some freedom in the choice of the "eigenvector" $^{(i)}X$; we fix the
choice in some way.) If ${\cal A}$ is a Banach algebra one can check that these series converge for sufficiently small
$\epsilon $.

Let us consider a matrix $\Xi$ having "eigenvectors" $^{(1)}X,...,^{(n)}X$ as its columns. It is easy to check that 
\begin {equation}
A\Xi =\Xi \Lambda
\end {equation}
where $\Lambda$ is a diagonal matrix with  entries $^{(1)}\lambda,...,^{(n)}\lambda$. The equation (6) was used in [2] to
obtain the information about noncommutative eigenvalues under the assumption that the matrix $\Xi$ is invertible. We are
working in the framework of perturbation theory, therefore $\Xi$ is always  invertible.  ($\Xi ^{-1}$ exists as a series with
respect to $\epsilon$, because the series for $\Xi$ starts with an  invertible matrix.) We can say that 
\begin {equation}
\Xi ^{-1}A\Xi =\Lambda
\end {equation}
It follows from (7) that 
\begin {equation}
\Xi^{-1}(\oint _{\Gamma}A(A-\zeta)^{-1}d\zeta)\Xi=\oint _{\Gamma} \Lambda(\Lambda-\zeta)^{-1}d\zeta
\end {equation}
Here $\Gamma$  is an arbitrary curve on ${\bf C}$ that does not contain $\kappa_1,...,\kappa_n$. The condition $\kappa_i\not
\in \Gamma$ permits us to say that $\Lambda -\zeta$ and therefore $A-\zeta$ are  invertible in the framework of perturbation
theory.

Let us assume now that $\Gamma$ is a closed curve and the domain $D$, bounded by $\Gamma$, contains only one of the points
$\kappa_1,...,\kappa_n$.  Then it follows from (8) that  
\begin {equation}
{\rm Tr} ^{(i)}\lambda =(2\pi i)^{-1}\oint _{\Gamma}{\rm Tr}A(A-\zeta)^{-1}d\zeta=(2\pi i)^{-1}\oint _{\Gamma}{\rm Tr}\zeta
(A-\zeta)^{-1}d\zeta
\end {equation}
(Here we assumed that $\kappa_i\in D$). As earlier Tr stands for an arbitrary trace on ${\cal A}$; we used the relation ${\rm
Tr} \Xi ^{-1} K \Xi ={\rm Tr}K$.)  Using the formula
$$(A-\zeta)^{-1}=(\epsilon B+P-\zeta)^{-1}=(1+(P-\zeta)^{-1}\epsilon B)^{-1}(P-\zeta)^{-1}$$

$$= (P-\zeta)^{-1}- (P-\zeta)^{-1}\epsilon B (P-\zeta)^{-1}+(P-\zeta)^{-1}  \epsilon  B  (P-\zeta)^{-1} \epsilon B
(P-\zeta)^{-1} +.....$$ 
we can easily obtain the perturbation series for ${\rm Tr}^{(i)}\lambda$. However, as in standard  perturbation theory (see
[6]) it is more convenient to rearrange this series using the relation
\begin {equation}
{\rm Tr}{d\over d\zeta}(BR(\zeta))^p=p{\rm Tr}R(\zeta)(BR(\zeta))^p.
\end {equation}
We introduced here the notation 
$$R(\zeta)=(P-\zeta)^{-1},$$
the relation (10) follows from $dR/d\zeta =R(\zeta)^2$.

We get the following perturbation series for the trace of noncommutative eigenvalue $^{(i)}\lambda$: 
\begin {equation}
{\rm Tr}^{(i)}\lambda =\kappa_i+{1\over 2\pi i}\sum _{p=1}\epsilon ^p{(-1)^p\over p}\oint _{\Gamma}{\rm
Tr}(BR(\zeta)...BR(\zeta))d\zeta
\end {equation}
where $\Gamma$ is a closed curve that encircles $\kappa _i$.
The above consideration can be generalized to obtain an expression for Tr$(^{(i)}\lambda)^r$. Namely, modifying slightly the
derivation of (9) and (11) we obtain
 $${\rm Tr}(^{(i)}\lambda)^r=   (2\pi i)^{-1}\oint _{\Gamma}{\rm Tr}A^r(A-\zeta)^{-1}d\zeta=(2\pi i)^{-1}\oint _{\Gamma}\zeta
^r{\rm Tr}(A-\zeta)^{-1}d\zeta $$
\begin {equation}
=\kappa _i^r+{1\over 2\pi i}\sum _{p=1}{\epsilon ^p(-1)^pr\over p}\oint _{\Gamma}\zeta ^{r-1}{\rm
Tr}BR(\zeta)...BR(\zeta)d\zeta .
\end {equation}
The formula we obtained is a generalization of well known formula (see for example [6], p.79). Recall that the trace in (12)
is an arbitrary linear functional on ${\cal A}$ that vanishes on all commutators, i.e. a linear functional on $\bar {{\cal
A}}={\cal A}/[{\cal A},{\cal A}]$. 
It is possible (and sometimes more convenient) to consider Tr in (12) as a natural map ${\cal A}\rightarrow\bar {{\cal A}}$. 

We can consider entries of the matrix $B$ as generators of free associative algebra. An element of free unital associative
algebra ${\cal F}$ with generators $e_1,...,e_n$ can be regarded as linear combination of expressions of the form $e_{\alpha
_1},...,e_{\alpha _p}$ , $p\geq 0$ (a linear combination of words with letters $e_i$). An element of   ${\cal F}/[{\cal
F},{\cal F}]$ can be considered as linear combination of cyclic words. (The group ${\bf Z}_p$ acts on the set of words of
length $p$ by means of cyclic permutations. Two words belonging to the same orbit are equal mod $[{\cal F},{\cal F}]$. This
means that every word $\omega$ is equal mod $[{\cal F},{\cal F}]$ to a cyclic word $\hat {\omega}$, i.e. to an average of all
words in its ${\bf Z}_p$-orbit. We identify  ${\cal F}/[{\cal F},{\cal F}]$ with the subspace of ${\cal F}$ spanned by all
cyclic words.) Using (12)  we can express Tr $^{(i)}\lambda$ as a linear combination of cyclic words. A cyclic word $\hat
{{\omega}}$ where $\omega=b_{\alpha_1,\beta_1}b_{\alpha_2,\beta_2}...b_{\alpha_p,\beta_p}$ enters this combination with
coefficient
\begin {equation}
c(\omega)=c_{\alpha_1,\beta_1,...,\alpha_p,\beta_p}={(-1)^p\epsilon ^p\over 2\pi i}\oint _{\Gamma}{\rm
Tr}(R_{\beta_1,\alpha_2}(\zeta)...R_{\beta_{p-1},\alpha_p}(\zeta) R_{\beta_p,\alpha_1}(\zeta))d\zeta.
\end {equation}
The analogous coefficient in the expression for Tr  $(^{(i)}\lambda)^r$ looks as follows:
\begin {equation}
c^{(r)}(\omega)={(c)}^r_{\alpha_1,\beta_1,...,\alpha_p,\beta_p}={(-1)^p\epsilon ^p r\over 2\pi i}\oint _{\Gamma}\zeta
^{r-1}{\rm Tr}(R_{\beta_1,\alpha_2}(\zeta)...R_{\beta_{p},\alpha_1}(\zeta) )d\zeta.
\end {equation}
Let us come back to the equation (1). For every solution $x$ of this equation we can construct a solution of eigenvalue
problem (2) with
\begin {equation}
A=\left ( \begin {array}{ccccc}
a_1 & .& . & . & a_n\\
1     &  &   &    &       \\
       & . &   &    &      \\
       &   & . &    &      \\
       &    &   & 1 &
\end {array}  \right )
\end {equation}
taking $\lambda =x$, $x_k=x^{n-k}$. This remark, that was used in [2], allows us to obtain information about solution of (1)
from the information about eigenvalue problem. We want to study (1) in the framework of perturbation theory; therefore we will
write (1) in the form 
\begin {equation}
x^n=\epsilon a_1x^{n-1}+...+ \epsilon a_n+1
\end {equation}
Then we have $n$  perturbative solutions that correspond to $n$ solutions of "unperturbed" equation. We can replace equation
(16) with the eigenvalue problem (3) with the matrix
\begin {equation}
A=\epsilon B+P\cdot 1=\epsilon \left ( \begin {array}{ccc} 
a_1 &  \cdots & a_n  \\
0     &              & 0   \\
        &             &       \\
0    &               &  0      \end {array}  \right )    + \left (  \begin {array} {ccccc}
0   & . &  . & . & 1  \\
1    &   &   &   &    \\
      & .  &   &    &   \\
      &    & .  &   &    \\
      &     &    & 1 &  0     
                                     \end {array}  \right )
\end {equation}
(Every perturbative solution $x(\epsilon)$ of (16) gives a perturbative solution of eigenvalue problem. We obtain one-to-one
correspondence because both problems have precisely $n$  perturbative solutions.)Now using (11) we get an explicit expression
for the traces of roots of (16). To apply (11) we should calculate 
$R(\zeta)=(P-\zeta)^{-1}$; it is convenient to express  it in the form  
$$R(\zeta)=(P-\zeta)^{-1}=-\zeta^{-1}(1-\zeta^{-1}P)^{-1}=\sum
_{\alpha=0}^{n-1} { \zeta^{n-\alpha-1}\over
1-\zeta^n}P^{\alpha}.$$
(We used that $P^n=1$). We see that 
\begin {equation}
\begin {array}{rcl}
R_{\alpha \beta}(\zeta)={\zeta ^{n-1-(\alpha-\beta)}\over 1-\zeta ^n} & {\rm if} &  \alpha  \geq  \beta    \\
R_{\alpha \beta}(\zeta)={\zeta ^{-1-(\alpha-\beta)}\over 1-\zeta ^n}  & {\rm if} &  \alpha <\beta 
\end {array}
\end {equation}
It is easy to verify using (13) that the cyclic word $\hat {\omega}$ where $\omega=a_{\alpha _1} ... a_{\alpha _p}$ enters the
perturbative expression for the trace of a root $x(\epsilon)$ of (16) with the coefficient
\begin {equation}
c(\omega)=c_{\alpha _1,...,\alpha _p}={(-1)^p\epsilon ^p\over 2\pi i}\oint _ {\Gamma}{\zeta ^{p(n-1)-\sum\alpha _i} \over
(1-\zeta ^n)^p}d\zeta .
\end {equation}
Corresponding coefficients for the trace of $x(\epsilon)^r$ can be obtained from (14):
\begin {equation}
c^{(r)}(\omega)=c_{\alpha _1,...,\alpha _p}^{(r)}={(-1)^p\epsilon ^p r\over 2\pi i}\oint _ {\Gamma}{\zeta
^{r+p(n-1)-\sum\alpha _i} \over (1-\zeta ^n)^p}d\zeta .
\end {equation}
{\it We obtain that} $c_{\alpha_1,...,\alpha_p}^{(r)}$  {\it is symmetric with respect to} $\alpha_1,...,\alpha _p$. This
statement was proved in [3] for $r=1$. It permits us to obtain a
perturbative expression for Tr $x(\epsilon)^r$ from the
solution of corresponding commutative problem.   

Namely, we can construct a linear map $\cal S$ of   commutative algebra of power series with respect to $a_1,...,a_n$ into
corresponding noncommutative object assigning to every monomial a sum of all words     corresponding to this monomial
multiplied by a normalization factor. (There exists a natural homomorphism $\pi$ of noncommutative  free algebra into
commutative polynomial algebra. The word $\omega$ corresponds to a monomial $\rho$ if $\pi (\omega)=\rho$. The normalization
factor is determined by the condition $\pi ({\cal S}(\rho))=\rho$.)

The following statement (proved in [3] for $r=1$) is an immediate
consequence of the symmetry of coefficients
$c_{\alpha_1,...,\alpha _p}^{(r)}$. 

{\it Let} $x_{comm}(\epsilon)$ {\it  be a perturbative solution of ordinary algebraic equation}  $x^n=\epsilon
(a_1x^{n-1}+...+a_n)+1$. {\it Then} Tr  $(x(\epsilon)^r)={\cal S}(x_{comm}(\epsilon)^r)$.

The integral 
\begin {equation}
\gamma _{n,p}^{\rho}=\oint _{\Gamma}{\sigma^{\rho}\over (1-\sigma^n)^p}d \sigma
\end {equation}
can be easily calculated [7]. One can use for example recursion formulas 
\begin {equation}
\gamma_{n,p}^{\rho}={n(1-p)+\rho+1 \over n(1-p)}\gamma_{n,p-1}^{\rho}
\end {equation}
and
\begin {equation}
\gamma_{n,p}^{\rho}={\rho+1-n \over \rho+1-np}\gamma_{n,p}^{\rho -n}.
\end {equation}
Using the notation (21) we can represent the coefficients $c_{\alpha_1,...,\alpha_p}^{(r)}$ in the form 
\begin {equation}
c_{\alpha _1,...,\alpha_p}^{(r)}={(-1)^p\epsilon^pr \over 2\pi i}\gamma_{n,p}^{r+p(n-1)-\sum \alpha _i}.
\end {equation}
We will use the formula (24) to derive some  statements about eigenvalues of  matrix $\epsilon B+P\cdot 1$ where 
\begin {equation}
P=\left ( \begin {array}{ccccc}
0 & .& . & . &  1 \\
1     &  &   &    &       \\
       & . &   &    &      \\
       &   & . &    &      \\
       &    &   & 1 &   0
\end {array}  \right )
\end {equation}
but $B$ is an arbitrary matrix. In this case using (20) and (21) we obtain the following expression for the coefficient
$c^{(r)}(\omega)=c_{\alpha _1,\beta _1,...,\alpha_p,\beta _p}^{(r)}$:
\begin {equation}
c_{\alpha _1,\beta _1,...,\alpha_p,\beta _p}^{(r)}={(-1)^p\epsilon^pr \over 2\pi i}\gamma_{n,p}^{r+\sum \alpha _i-\sum \beta
_i-p+Kn} 
\end {equation}
where $K$ is the number of indices $i$ obeying  $\beta _i\geq \alpha_{i+1}$ (we identify $\alpha _{p+1}$ 
 with $\alpha _1$). Let $\omega ^{\prime}$ be a word   corresponding to the same monomial as $\omega =b_{\alpha _1\beta
_1}...b_{\alpha_p\beta _p}$ (i.e. $\omega ^{\prime}$  is obtained from $\omega$ by means of permutation of factors
$b_{\alpha_i\beta _i}$). It is clear that the expression for $c(\omega ^{\prime})$ is almost identical to the expression  for
$c(\omega)$; only $K$ changes. We can use (23) to find $c(\omega^{\prime})/c(\omega)$; if $K(\omega^{\prime})-K(\omega)=s \geq
0$ we obtain
\begin {equation}
{c^{(r)}(\omega^{\prime})\over c^{(r)}(\omega)}={\rho +1-n\over \rho+1-np}\cdot {\rho +1-2n\over \rho+
1-n(p+1)}\cdot \cdot\cdot\cdot {\rho +1-ns\over \rho+1-n(p+s)}      
\end {equation}
where $\rho=\sum \alpha_i-\sum \beta_i-p+r$.
It is easy to check that using (27) we can find the perturbative expression  for the trace of noncommutative eigenvalue if we
know   the solution of corresponding commutative problem. However, it seems that the explicit   expression (26) is more
convenient.

{\bf Acknowledgments}

I am indebted to D.Brace, A.Chorin,I.Gelfand, M. Kapranov, B. Morariu,V.
Retakh and B.Zumino for useful discussions.

\centerline {\bf References.}

  1. Fuchs, D. ,Schwarz, A., {\it Matrix Vieta Theorem}, Lie groups and Lie 
algebras, E. B. Dynkin Seminar, (1996), 15-22,
Amer. Math. Soc. Transl. Ser. 2, 169

  2. Connes, A., Schwarz, A., {\it Matrix Vieta Theorem Revisited},
Lett. Math. Phys.  39 (1997), 349-353

     3.   Aschieri, P.,   Brace, D., Morariu, B., Zumino, B., {\it  Proof of a Symmetrized Trace Conjecture for the Abelian
Born-Infeld Lagrangian},  hep-th/0003228

     4.   Brace, D., Morariu, B., Zumino, B., {\it Duality Invariant 
Born-Infeld Theory}, Yuri Golfand memorial volume; hep-th/9905218

     5.      Aschieri,   P.,  Brace, D., Morariu, B., Zumino, B., 
{\it Nonlinear Self-Duality in Even Dimensions}, hep-th/9909021, to appear in Nucl. Phys. B

     6.   Kato,  T.  {\it Perturbation Theory  for Linear Operators},
Springer, 1995

     7.  Gradshteyn, I.  Ryzhik, I.,  {\it Table of integrals, series, and products}, Acad. Press, 1994.

     8. Etingof, P., Gelfand, I. and Retakh, V.,
{\it  Factorization of Differential Operators, 
Quasideterminants, and Nonabelian Toda Field Equations}, 
 Math. Research Letters, 4 ( 1997), q-alg/9701008

9. Gelfand, I., Krob, D., Lascoux, A., Retakh, V. 
and Thibon, J-Y., {\it Noncommutative Symmetric 
Functions}, Advances in Math., 112 (1995), 218-348, hep-th/9407124

10. Gelfand, I., and Retakh, V., {\it 
Determinants of matrices over noncommutative rings},
 Funct.An. Appl., 25:2 (1991), 91-102

11. Gelfand, I., and Retakh, V., {\it  A theory of noncommutative 
determinants and characteristic functions 
of graphs}, Funct.An. Appl., 26:4 (1992), 1-20

12. Gelfand, I., and Retakh, V., {\it  Noncommutative 
Vieta theorem and symmetric functions}, 
in: Gelfand Math. Seminars 1993-95, Birkhauser, 
Boston (1996), q-alg/9507010

13. Gelfand, I., and Retakh, V., {\it A theory of 
noncommutative determinants and characteristic functions
of graphs.I}, in: Publ. LACIM, UQAM, 14 (1993), 1-26

14. Gelfand, I., and Retakh, V., {\it  Quasideterminants, I}, 
Selecta Math., 3 (1997), 417-546, q-alg/9705026

\end {document}